\documentclass[letterpaper]{iopart}
\usepackage{cite}
\usepackage{amsmath, amssymb}
\usepackage{amsbsy, latexsym, amsfonts, amsthm}
\usepackage{mathtools}
\usepackage{color}

\newcommand{\dummy}{\rule{0mm}{0mm}}

\begin{document}

\title{Addendum: EPRL/FK Asymptotics and the Flatness Problem}
\author{J. Engle${}^{1,3}$, W. Kaminski${}^{2,3}$, J. Oliveira${}^4$}
\address{${}^1$\begin{minipage}[t]{4in}
Department of Physics, Florida Atlantic University,\\
\dummy \quad 777 Glades Road, Boca Raton, FL 33431 USA\end{minipage}}
\address{${}^2$\begin{minipage}[t]{4in}
Instytut Fizyki Teoretycznej, Uniwersytet Warszawski,\\
\dummy \quad ul. Ho\.za 69, 00-681 Warsaw, Poland\end{minipage}}
\address{${}^3$\begin{minipage}[t]{4in}
Institut f\"ur Quantengravitation, Department Physik Universit\"at Erlangen,\\ 
\dummy \quad Staudtstrasse 7, D-91058 Erlangen, Germany\end{minipage}}
\address{${}^4$\begin{minipage}[t]{4in}
School of Mathematical Sciences, University of Nottingham, \\
\dummy \quad University Park, Nottingham NG7 2RD, UK\end{minipage}}
\ead{\begin{minipage}[t]{3in}jonathan.engle@fau.edu, wojciech.kaminski@fuw.edu.pl,
josercamoes@gmail.com \end{minipage}}


\begin{abstract}
We show that, when an approximation used in this prior work is removed, the resulting improved calculation yields an alternative derivation, in the particular case studied, of the accidental curvature constraint of Hellmann and Kaminski. The result is at the same time extended to apply to almost all non-degenerate Regge-like boundary data and a broad class of face amplitudes.
This resolves a tension in the literature. 
\end{abstract}

Spin-foam models offer a space-time covariant, path integral definition of dynamics for loop quantum gravity, a canonical approach to quantizing general relativity restricted by Einstein's principle of general covariance, which is equivalent to \textit{background independence}. At present, the spin-foam model which has passed the most tests of viability, and on which the most work has been done, is the EPRL/FK model, which comes in both Lorentzian and Euclidean versions
\cite{elpr2007, fk2007}. The work \cite{oliveira2017} investigates the classical limit of the transition amplitude predicted by the Euclidean version of this model for the simplest triangulation allowing for space-time curvature. That is, it investigates the transition amplitude for the simplest triangulation with an internal triangle, in the limit of large boundary quantum numbers. 

The boundary states considered are Livine-Speziale states \cite{ls2007}, which are peaked on the three dimensional intrinsic boundary geometry. For the simple triangulation considered, such boundary data, classically, uniquely determines the area of, and deficit angle around, the one interior triangle.  Two different sets of boundary data are considered in \cite{oliveira2017}: One in which the classically determined deficit angle is zero (flat), and one in which it non-zero (curved).
In evaluating the transition amplitudes for these boundary states, this work goes beyond the earlier work \cite{mp2011b} in that each term in the discrete sum over the internal spin, determining the area of the internal triangle, is analyzed in more detail.
%
%
The work \cite{oliveira2017} concludes that the transition amplitudes for 
both of the boundary states considered are not exponentially suppressed, 
in agreement with the earlier work \cite{mp2011b}.

This result seems to contradict the works \cite{hk2012, perini2012, hk2013, han2013}, which 
conclude, for both the Euclidean and Lorentzian EPRL models, that, in order for the transition amplitude to not be exponentially suppressed in the classical limit, every internal deficit angle $\Theta$ must satisfy an \textit{accidental curvature constraint}
%
%
\begin{equation}
\label{accidentalcurvature}
\gamma \Theta = 0 \text{ mod }4\pi.
\end{equation}
This is a version of the so-called `flatness problem' originally derived for Euclidean EPRL and SU(2) BF theory \cite{bonzom2009},
but more precise in that it considers the discrete nature of the sums over internal spins. 
%
%
The data for the \textit{curved} triangulation considered by \cite{oliveira2017} is that of the three 4-simplices adjoining a common triangle in the boundary of a regular 5-simplex. Each of these 4-simplices is regular, whence they all  have dihedral angle $\arccos(1/4)$, yielding the internal deficit angle
\begin{displaymath}
\Theta = 2\pi - 3 \arccos(1/4) \approx 2.32884 .
\end{displaymath}
The value of $\gamma$ used in the calculation of the corresponding transition amplitude is $\gamma = \sqrt{3}/2$, so that
\begin{displaymath}
\frac{\gamma \Theta}{4\pi} \approx 0.16049,
\end{displaymath}
which is not an integer. Equation (\ref{accidentalcurvature}) is thus violated, so that the results of \cite{hk2012,hk2013, han2013} imply that the amplitude should be suppressed.

How can this be consistent with the conclusion of \cite{oliveira2017} that it is not suppressed? 
As we shall show below, the calculation in \cite{oliveira2017} makes an assumption, which, when removed, leads exactly to the constraint (\ref{accidentalcurvature}), thereby providing a new derivation of the constraint in this case. 
We also take the opportunity, at the end of this Addendum, to make the argument in the earlier paper \cite{mp2011b} more precise to show exactly where it falls short. This relieves the tension between \cite{oliveira2017, mp2011b} and the works \cite{hk2012, perini2012, hk2013, han2013}.

We would like to emphasize that the accidental curvature constraint (\ref{accidentalcurvature}) is not necessarily a problem for the EPRL model, but may be an artifact of attempting to take the classical limit of the theory prior to taking the refinement limit of the theory, or, equivalently, prior to summing over all triangulations, a step integral to the theory's complete definition \cite{rr1996,rovelli1998}.
This is related to the viewpoint first expressed and developed by Han in \cite{han2013, han2017}, that the correct classical limit involves not only the limit of large quantum numbers, but also the refinement limit. 
In this combined limit, the spin-foam sum includes a family of discrete geometries approximating any desired curved smooth geometry, in which the deficit angles nevertheless approach zero and so approach satisfying (\ref{accidentalcurvature}) arbitrarily well.

Furthermore, the recent numerical work of \cite{dgs2020, gozzini2021talk} (in the closely related $SU(2)$ BF and Lorentzian EPRL models), going up to boundary spin values of 30, could not see the exponential suppression from the accidental curvature constraint. 
This suppression could only be seen by increasing the expected deficit angle and going up to boundary spin values of 40 thanks to an improved code \cite{dghs2021private, gozzini2021, dgs2020}.
%
%
%
That is to say: numerically it is found that the exponential suppression from the accidental curvature constraint is much slower than that from the orientation and closure constraints essential in recovering Regge geometries. This yields a range of boundary spins high enough that orientation and closure are imposed, but low enough that the accidental curvature constraint is not --- a range in which correct physics can be modeled, even for a fixed triangulation. 
This is similar to what is found in \cite{adh2020, adh2020a}, where, for a simpler but similar quantum gravity model, good semi-classical behavior is shown even on a fixed triangulation for certain ranges of spins and parameters.


\section*{Calculation}

The transition amplitude under consideration is for Regge-like, non-degenerate boundary data in which the boundary spins are scaled by a factor $\lambda$.
\cite{oliveira2017} casts this transition amplitude in the 
form
%
%
\begin{align}
\begin{split}
\label{amp}
Z(\lambda) &= \sum_{j' \in [j'_\text{min}, j'_\text{max}] \cap \frac{\mathbb{Z}}{2\lambda}}  \mu(\lambda j') \int_{\mathcal{M}_n}
e^{i \lambda S(\vec{x},j')} \nu(\vec{x}) d^n \vec{x}\\
&= \sum_{j' \in [j'_\text{min}, j'_\text{max}]}  \chi_{\frac{\mathbb{Z}}{2\lambda}}(j') \mu(\lambda j') \int_{\mathcal{M}_n}
e^{i \lambda S(\vec{x},j')} \nu(\vec{x}) d^n \vec{x}
\end{split}
\end{align}
where $\mathcal{M}_n$ is a manifold of dimension 
$n =78$\footnote{consisting of the 4 non-gauge-fixed interior $Spin(4)$ parallel transports in each of the three 4-simplices, and the $S^2$ normal to the interior triangle in the frame of each of the three interior tetrahedra}
with coordinates 
$\vec{x} \in \mathbb{R}^n$ defined almost everywhere, $S(\vec{x},j')$ is complex valued, $\nu(\vec{x})$ real valued, 
$0\le j'_{min}\le j'_{max}$ with $j'_{min}$ and $j'_{max}$ determined by the fixed boundary data via 
triangle inequalities, and $\chi_S$ denotes the 
characteristic function of $S$.
$j = \lambda j' \in \frac{\mathbb{N}}{2}$ is the quantum number determining the area of the internal triangle, which may be taken to be $8\pi \gamma \hbar G \lambda j'$, where $\gamma$ is the Barbero-Immirzi parameter.
%
%
$\mu(j)$ is the face amplitude, usually taken to be $2j+1$ from the arguments of \cite{brr2010}.
We keep, however, $\mu(j)$ general: The analysis below will apply to any choice of $\mu(j)$ admitting an asymptotic expansion in powers of $1/j$.
Denote the leading order term in this expansion by 
$\mu_o j^p$ with $p \in \mathbb{Z}$.

The goal is to obtain an asymptotic expression for $Z(\lambda)$ in the limit in which the coherent boundary state becomes more classical: specifically, the limit of large quantum numbers $\lambda \rightarrow \infty$.
To this end, \cite{oliveira2017} takes the asymptotic limit of the summand as $\lambda \rightarrow \infty$ for each $j'$. 
As noted earlier, for this simple 
triangulation, the fixed boundary data, together with the Euclidean geometry of each 4-simplex,
determines the area of the internal 
triangle to be some unique value 
$8\pi \gamma \hbar G \lambda j'_o$.
For simplicity, assume that $j'_o \neq 0$, as for the boundary data used in \cite{oliveira2017}.
As a consequence, the complex spin-foam action $S(\vec{x},j')$ in each of the integrals in (\ref{amp})
has a critical point $\vec{x}_o$ with respect to its dependence on $\vec{x}$ for only the single value $j'_o \in [j'_{min}, j'_{max}]$ of $j'$.
\footnote{For a real action, the implicit function theorem would guarantee that there exists critical points in some neighborhood of $j'_o$; but the action in this case is complex, and in fact there are no critical points in any neighborhood of $j'_o$.}
Hence, for each $j'\neq j'_o$ in the sum, the usual stationary phase theorem (7.7.5 in \cite{hormander1983}) 
implies exponential decay of the corresponding term, 
in consistency with the classical theory. 
In order to determine whether the \textit{full sum} is exponentially suppressed, however, one needs more information about these exponentially suppressed terms.
We first note that, 
if $j'_{min} = 0$, then, because $j'_o \neq 0$,
the $j'=0$ term will be exponentially suppressed and can be dropped from the expression for the asymptotic limit.
\newcommand{\sbrack}[1]{\,{}^*\![#1]}
To aide in denoting this, we let $\sbrack{a,b}$
denote the interval $(a,b]$ if $j'_{min} = 0$
and $[a,b]$ otherwise.
For the remaining terms in the sum, 
the $\lambda \rightarrow \infty$ 
asymptotic limits of the face factors
$\mu(\lambda j')$ are $\mu_o \cdot (\lambda j')^p$,
so that
\begin{align}
\label{muasym}
Z(\lambda) \sim \mu_o \lambda^p \sum_{j' \in \sbrack{j'_\text{min}, j'_\text{max}}} \chi_{\frac{\mathbb{Z}}{2\lambda}}(j')  \int_{\mathcal{M}_n}
e^{i \lambda S(\vec{x},j')} (j')^p \nu(\vec{x}) d^n \vec{x} .
\end{align}
\cite{oliveira2017} then uses an elegant extension of 
stationary phase, namely theorem 7.7.12 in \cite{hormander1983},
which provides a non-trivial exact asymptotic expression of the above integral
for each $j'$ in a neighborhood of $j'_o$. 
Before applying this theorem, \cite{oliveira2017}
approximates the spin-foam action $S(\vec{x},j')$ by its second order Taylor polynomial in the $x_i$'s and $j'$ in a neighborhood of the one critical point
$(\vec{x}_o, j'_o)$.  
With this approximation, the application of theorem 7.7.12 is equivalent to  completing the squares in the $x_i$ variables, 
and using the fact that the widths of the resulting Gaussians go to zero in the
$\lambda \rightarrow \infty$ limit to evaluate the integrals as Gaussian integrals. 
One obtains
\begin{align}
\label{Zasym}
Z(\lambda) &\approx 
C \lambda^{p+n/2}
\sum_{j' \in \sbrack{j'_\text{min}, j'_\text{max}} \cap \frac{\mathbb{Z}}{2\lambda}} 
\exp \lambda i\left(S_o^R + \gamma \Theta (j'-j'_o) + a (j'-j'_o)^2\right)
\end{align}
where $S^R_o$ is the value of the Regge action evaluated on the geometry fixed by the boundary data, $\Theta$ is, upto a sign, the deficit angle on the interior triangle in this geometry, and
\begin{align*}
C&:=  \frac{(2\pi i)^{n/2}\mu_o (j'_o)^p \nu(\vec{x}_o)}{\sqrt{\det H}}\\
a&:= \frac{-1}{2} (K_i H^{ij} K_j) 
\end{align*}
with
\begin{align*}
 H_{ij} &:= \frac{\partial^2 S(\vec{x},j')}{\partial x^i \partial x^j}(\vec{x}_o, j'_o), \\
 K_i &:= \frac{\partial^2 S(\vec{x}, j')}{\partial x^i \partial j'}(\vec{x}_o, j'_o),
\end{align*}
$\det H$ the determinant of $H_{ij}$, and $H^{ij}$ its inverse.

At this point, \cite{oliveira2017} assumes that only the real part of the argument of the exponential in (\ref{Zasym}) is relevant in determining whether $Z(\lambda)$ is exponentially suppressed as $\lambda \rightarrow \infty$, and so the imaginary part of the argument is dropped in equation (72) and subsequent equations of \cite{oliveira2017}, concluding in the end that the amplitude is not suppressed even for the boundary data forcing a curved interior. We argue now that this assumption and conclusion are incorrect: We retain the imaginary part of the argument of the exponential in (\ref{Zasym}) and show that it leads to exponential suppression in the case considered, and more generally leads to exponential suppression unless (\ref{accidentalcurvature}) is satisfied.

Changing the variable $j'$ in favor of $m:= 2\lambda (j'-j'_o)$, (\ref{Zasym}) becomes
\begin{align}
\label{Zasymm}
Z(\lambda) &\approx 
C \lambda^{p+n/2} e^{i\lambda S_o^R}  \hspace{-1cm}
\sum_{m \in \sbrack{-2\lambda (j'_o-j'_\text{min}), 2\lambda(j'_\text{max}- j'_o)} \cap \mathbb{Z}} \hspace{-0.5cm}
\exp i \left( \frac{\gamma \Theta m}{2} + \frac{a m^2}{4\lambda}\right) .
\end{align}
Now, the imaginary part of the spin-foam action $S(\vec{x},j')$,
with the conventions here, is always non-negative and is equal to zero at all critical points. $\Im S(\vec{x}, j')$ is thus minimized at $(\vec{x}_o, j'_o)$, so that its second derivative there, along any curve, is non-negative. Application of this to curves straight in the coordinate system $(\vec{x}, j')$, combined with the symmetry of the second derivative matrix $M$
\begin{align*}
    M := \left(\begin{array}{cc} 0 & K^T \\ K & H \end{array}\right),
\end{align*}
implies that the (component-wise) imaginary part of $M$ is positive semi-definite. (Note that since $S(\vec{x}, j')$ is linear in $j'$ \cite{oliveira2017, bdfgh2009, hz2011}, the $j'$-$j'$ component of $M$ is zero.) Thus, defining 
\begin{align*}
    J:= \left(\begin{array}{c} 1 \\ -H^{-1} K \end{array}\right),
\end{align*}
one can check that $a = \frac{1}{2} J^\dagger M J$, 
and we have
\begin{align*}
    \Im a &= \frac{1}{4i}\left(J^\dagger M J - \overline{J^\dagger M J}\right) 
    = \frac{1}{4i}\left(J^\dagger M J - J^\dagger M^\dagger J\right) = J^\dagger \frac{1}{4i}\left(M - M^\dagger \right) J \\
    &= J^\dagger \frac{1}{4i}\left(M - \overline{M} \right) J 
    = \frac{1}{2} J^\dagger (\Im M) J \\
    & \ge 0 
\end{align*}
always.
In order for the quadratic approximation in (\ref{Zasym}) to correctly reflect the qualitative fact that the integral in (\ref{muasym}) is suppressed for $j'\neq j'_o$, it is necessary to assume that the above inequality is furthermore \textit{strict}, $\Im a > 0$.
This happens for every case explicitly checked in \cite{oliveira2017}\footnote{From equation (99) in \cite{oliveira2017} and 
$\Im a = A/8$, $\Im a \approx 0.82753>0$ and $\Im a \approx 1.8049> 0$ respectively for the curved and flat cases considered there.}.

The summand in (\ref{Zasymm}) is then a Gaussian function of $m$ peaked about $m=0$
with standard deviation $\sqrt{2\lambda/\Im a}$. 
Since the width of the summation domain grows faster than the standard deviation, 
the lower and upper limits of the sum may be dropped in the 
$\lambda \rightarrow \infty$ limit, leaving 
\begin{align}
\label{Zexp}
Z(\lambda) &\approx 
C \lambda^{p+n/2} e^{i\lambda S_o^R} \sum_{m \in \mathbb{Z}} 
\exp i\left(\frac{\gamma \Theta m}{2} + \frac{a m^2}{4\lambda}\right) .
\end{align}
Setting $x:= 2m \pi$, let $f(x)$ denote the summand in 
(\ref{Zexp}), so that its Fourier transform is
\begin{align*}
\tilde{f}(k) := \frac{1}{\sqrt{2\pi}} \int_{-\infty}^{\infty} \exp i\left(\frac{\gamma \Theta x}{4\pi} + \frac{a x^2}{\lambda(4\pi)^2} - kx\right) dx
= \pi \sqrt{\frac{8 \lambda}{ia}} \exp \frac{-i\lambda}{4a}
\left(4\pi k - \gamma \Theta\right)^2.
\end{align*}
Poisson resummation of (\ref{Zexp}) then yields
\begin{align}
\label{poisson}
Z(\lambda) &\approx 
C' \lambda^{(2p+n+1)/2} e^{i\lambda S_R^o}
\sum_{k=-\infty}^{\infty} \exp \frac{-i\lambda}{4 a}
\left(4\pi k - \gamma \Theta\right)^2
\end{align}
where $C'$ is independent of $\lambda$.

\textit{Suppose $\gamma \Theta/(4\pi) \not\in \mathbb{Z}$}.
Let $k_-$ and $k_+$ be the integers directly above and below 
$\gamma \Theta/(4\pi)$.  We then  have
\begin{align*}
&|Z(\lambda)| \le |C'| \lambda^{(2p+n+1)/2} 
\sum_{k=-\infty}^{\infty} \left|\exp \frac{-i\lambda}{4 a}
\left(4\pi k - \gamma \Theta \right)^2 \right|\\
&=
|C'| \lambda^{(2p+n+1)/2} 
\sum_{k=-\infty}^{\infty} \exp \frac{-\lambda \Im a}{4 |a|^2}
\left(4\pi k - \gamma \Theta\right)^2 \\
&= 
|C'| \lambda^{(2p+n+1)/2}\Bigg(
\sum_{k=-\infty}^{k_--1} \exp \frac{-\lambda \Im a}{4 |a|^2}
\left(4\pi k - \gamma \Theta \right)^2 
+ \exp \frac{-\lambda \Im a}{4 |a|^2}
\left(4\pi k_- -\gamma \Theta\right)^2 \\
& \dummy \qquad + \exp \frac{-\lambda \Im a}{4 |a|^2}
\left(4\pi k_+ - \gamma \Theta\right)^2 +
\sum_{k=k_++1}^{\infty} \exp \frac{-\lambda \Im a}{4 |a|^2}
\left(4\pi k - \gamma \Theta\right)^2\Bigg) \\
&\le |C'| \lambda^{(2p+n+1)/2}\Bigg(
\int_{\infty}^{k_-} \exp \frac{-\lambda \Im a}{4 |a|^2}
\left(4\pi k - \gamma \Theta\right)^2 dk
+ \exp \frac{-\lambda \Im a}{4 |a|^2}
\left(4\pi k_- - \gamma \Theta\right)^2 \\
& \dummy \qquad + \exp \frac{-\lambda \Im a}{4 |a|^2}
\left(4\pi k_+ - \gamma \Theta\right)^2 +
\int_{k_+}^{\infty} \exp \frac{-\lambda \Im a}{4 |a|^2}
\left(4\pi k - \gamma \Theta\right)^2 dk \Bigg)\\
&=|C'| \lambda^{(2p+n+1)/2}\Bigg(
\frac{|a|}{4\sqrt{\pi \lambda \Im a}}
\mathrm{erfc}\left(\frac{\sqrt{\lambda \Im a}}{2|a|}(\gamma \Theta - 4\pi k_-)\right)
+ \exp \frac{-\lambda \Im a}{4 |a|^2}
\left(4\pi k_- - \gamma \Theta\right)^2 \\
& \dummy \qquad + \exp \frac{-\lambda \Im a}{4 |a|^2}
\left(4\pi k_+ - \gamma \Theta\right)^2 +
\frac{|a|}{4\sqrt{\pi \lambda \Im a}}
\mathrm{erfc}\left(\frac{\sqrt{\lambda \Im a}}{2|a|}(4\pi k_+ - \gamma\Theta)\right) \Bigg) \\
&\sim |C'| \lambda^{(2p+n+1)/2}\!\Bigg(\!
\frac{|a|^2}{\pi \lambda \Im a (\gamma\Theta\! -\! 4\pi k_-\!)}
\exp \frac{-\lambda \Im a}{4 |a|^2}\!
\left(\!4\pi k_- \!\!\!-\! \gamma \Theta\!\right)^2 
+ \exp \frac{-\lambda \Im a}{4 |a|^2}\!
\left(\!4\pi k_- \!\!\!-\! \gamma \Theta\!\right)^2 \\ 
& \dummy \qquad + \exp \frac{-\lambda \Im a}{4 |a|^2}
\left(4\pi k_+ - \gamma \Theta\right)^2 +
\frac{|a|^2}{\pi \lambda \Im a (4\pi k_+ - \gamma\Theta)}
\exp \frac{-\lambda \Im a}{4 |a|^2}
\left(4\pi k_+ - \gamma\Theta\right)^2\Bigg) \\
&\sim |C'| \lambda^{(2p+n+1)/2}\Bigg(
\exp \frac{-\lambda \Im a}{4 |a|^2}
\left(4\pi k_- - \gamma \Theta\right)^2 
+ \exp \frac{-\lambda \Im a}{4 |a|^2}
\left(4\pi k_+ - \gamma \Theta\right)^2\Bigg) 
\end{align*}
Since $\Im a>0$, both of these terms are
exponentially suppressed as $\lambda \rightarrow \infty$.

Thus, in the $\lambda \rightarrow \infty$
limit, if $\gamma \Theta/(4\pi) \not\in \mathbb{Z}$, the amplitude is exponentially suppressed.  By contrast, if $\gamma \Theta/(4\pi) = k_o$ for 
some integer $k_o$, similar methods show that the sum of all of the terms in the sum (\ref{poisson})
\textit{with $k=k_o$ removed} is exponentially suppressed, so that the amplitude is asymptotically equal to the $k=k_o$ term.
That is, the amplitude is exponentially suppressed if and only if the accidental curvature constraint (\ref{accidentalcurvature}) is satisfied.

\subsection*{Remark on Magliaro-Perini argument}

The exact expression (\ref{amp}) for the amplitude enables one to 
construct a more precise version of the argument in \cite{mp2011b}, which
argued that the amplitude is never exponentially suppressed for any $\Theta$. 
One thereby can see
where the argument fails, so that there is no contradiction with the result presented above. Specifically, similar to \cite{mp2011b}, we can write
\begin{align}
\label{Zsep}
Z(\lambda) = \left(\chi_{\frac{\mathbb{Z}}{2\lambda}}(j'_o) \mu(\lambda j'_o) \int_{\mathcal{M}_n}
e^{i \lambda S(\vec{x},j'_o)} \nu(\vec{x}) d^n \vec{x}\right)
+ R(\lambda)
\end{align}
where the remainder $R(\lambda)$ is given by
\begin{align}
\label{rem}
R(\lambda) &= 
 \sum_{j' \in ([j'_{min},j'_{max}]\setminus\{j'_o\})\cap\frac{\mathbb{Z}}{2\lambda}} R_{j'}(\lambda)
\end{align}
with
\begin{align}
\label{remterm}
R_{j'}(\lambda) &= \mu(\lambda j') \int_{\mathcal{M}_n}
e^{i \lambda S(\vec{x},j')} \nu(\vec{x}) d^n \vec{x} .
\end{align}
The first term in (\ref{Zsep}) is an integral with a critical point, and 
so does not exponentially decay, but is asymptotically given by 
\begin{align}
\label{nodecay}
C \lambda^{n/2}
\chi_{\frac{\mathbb{Z}}{2\lambda}}(j'_o)
\mu(\lambda j'_o) \exp \lambda i S_o^R .
\end{align}
The strategy of \cite{mp2011b}
is to show that the remainder $R(\lambda)$ \textit{does} exponentially decay,
so that the total amplitude (\ref{Zsep}) is asymptotically given by 
(\ref{nodecay}) and hence does not exponentially decay.
Each term in (\ref{rem}) has $j'\neq j'_o$, 
so that there is no critical point in any of the corresponding integrals
(\ref{remterm}). Consequently, each integral (\ref{remterm}) is 
indeed exponentially suppressed in the sense
that, for each $\ell \in \mathbb{N}$, there exists 
\newcommand{\constjl}{C_{j'\!,\ell}}
$\constjl \in \mathbb{R}^+$ such that 
\begin{align*}
|R_{j'}(\lambda)| < \constjl \, \lambda^{-\ell}.
\end{align*}
Equation (\ref{rem}) and the triangle inequality then imply
\begin{align}
\label{rembound}
|R(\lambda)| < 
C_\ell(\lambda) \lambda^{-\ell}
\end{align}
with
\begin{align}
\label{const}
C_\ell(\lambda) := \sum_{j' \in ([j'_{min},j'_{max}]\setminus\{j'_o\}) \cap \frac{\mathbb{Z}}{2\lambda}} \constjl . 
\end{align}
The essential argument of \cite{mp2011b} (translated into the present language) is to first set
\begin{align*}
A_\ell := \sup_{j' \neq j'_o} \constjl.
\end{align*}
Then, because, for any finite $\lambda$, the number of terms in (\ref{const})
is finite and bounded by $B_\ell \lambda$ for some $B_\ell$, we have
\begin{align*}
C_\ell(\lambda) < A_\ell B_\ell \lambda ,
\end{align*}
so that (\ref{rembound}) yields
\begin{align}
\label{remboundb}
|R(\lambda)| < A_\ell B_\ell  \lambda^{-\ell + 1} .
\end{align}
What is missed in this argument is that, in the limit 
$\lambda \rightarrow \infty$, the sum (\ref{const})
includes $j'$ arbitrarily close to the critical point $j'_o$, 
in which limit $\constjl$ must diverge for sufficiently large $\ell$.
That is, for sufficiently large $\ell$, 
the set of constants $\{\constjl\}$ appearing in the sum (\ref{const})
is not bounded, whence $A_\ell$ is infinite and cannot be used 
in (\ref{remboundb}) to imply exponential suppression of $R(\lambda)$.

\section*{Acknowledgements}
JE was supported in part by NSF grant 1806290. WK is supported by Polish National Science Centre grant Sheng 1 2018/30/Q/ST2/00811. 
JE and WK are also grateful for the hospitality of the Institut f\"ur Quantengravitation, Erlangen, where this clarification arose in discussions after a seminar by WK.

\vspace{1.5em}


\end{document}